\documentstyle[aps,prb,epsf]{revtex}
\begin{document}
\draft
\title{Cumulant approach to weakly doped antiferromagnets}
\author{Matthias Vojta and Klaus W. Becker}
\address{Institut f\"{u}r Theoretische Physik,
Technische Universit\"{a}t Dresden, D-01062 Dresden, \\ Germany}
\maketitle
\begin{abstract}
We present a new approach to static and dynamical properties of
holes and spins in weakly doped
antiferromagnets in two dimensions.
The calculations are based on a recently introduced cumulant approach
to ground--state properties of correlated electronic systems.
The present method allows to evaluate hole and spin--wave dispersion
relations by considering hole or spin excitations of the ground
state. Usually, these dispersions
are found from time--dependent correlation functions.
To demonstrate the ability of the approach we first
derive the dispersion relation for the lowest single
hole excitation at half--filling. However,
the main purpose of this paper is to focus on the
mutual influence of mobile holes and spin waves in the weakly
doped system. It is shown that low-energy spin excitations strongly admix
to the ground--state. The coupling of spin waves and holes
leads to a strong suppression of the staggered
magnetization which can not be explained by a simple rigid--band picture
for the hole quasiparticles.
Also the experimentally observed doping dependence of
the spin--wave excitation energies can be understood within our formalism.
\\
\end{abstract}

\pacs{PACS codes: 74.25.Ha, 75.10.Jm, 75.30.Ds, 75.50.Ee}

\section{Introduction}
The physical properties of high--temperature superconductors are strongly
influenced by electronic correlations. The investigation of strongly
correlated electronic systems has therefore become a major topic for the
understanding of the high--$T_c$ materials. The undoped compounds
are antiferromagnetic Mott--Hubbard insulators. Neutron scattering
experiments\cite{RossMig91,RossMig93} show the existence of spin--wave
excitations which can be described by conventional spin--wave theory for the
isotropic spin $S={1 \over 2}$ Heisenberg model on a square lattice
\cite{Anderson52}. The doped materials show a strong dependence
of the magnetic properties on the hole concentration $\delta$
in the $\rm CuO_2$ planes. With increasing hole concentration both
the N\'{e}el temperature and the staggered magnetization
decrease and vanish at a critical hole
concentration $\delta_c$ of a few percent before the system
becomes paramagnetic and metallic (or superconducting at sufficiently low
temperatures). Upon doping also the spin--wave velocity decreases
and vanishes at approximately the same hole
concentration $\delta_c$. At the same time the
long--wavelength spin-wave modes become overdamped.

The essential aspects of the low--energetic
electronic degrees of freedom of the $\rm CuO_2$ planes
are by now believed to be well described by the two--dimensional $t$-$J$
model\cite{Anderson87,Zhang88}:
\begin{equation}
H\, =\, - t \sum_{\langle ij\rangle \sigma}
      (\hat c^\dagger_{i\sigma} \hat c_{j\sigma} +
       \hat c^\dagger_{j\sigma} \hat c_{i\sigma})
    + J \sum_{\langle ij\rangle} \ ({\bf S}_i {\bf S}_j - {{n_i n_j} \over 4} )
\,.
\label{TJ_MOD}
\end{equation}
Here, ${\bf S}_i$ is the electronic spin operator and $n_i$ the electron
number operator at site $i$. The symbol $\langle ij \rangle$ refers to
a summation over pairs of nearest neighbors.
In the following we denote the two antiferromagnetic
sublattices by $\uparrow$ and $\downarrow$.
In all sums we shall use $i \in \uparrow$ and $j \in \downarrow$.
Note that the Hamiltonian (\ref{TJ_MOD}) is defined in the subspace of the
unitary space without double occupations of sites. The
electronic creation operators $\hat c_{i \sigma}^{\dagger}$ are not usual
fermion operators but rather exclude double occupancies:
\begin{equation}
\hat c^{\dagger}_{i\sigma} = c^{\dagger}_{i\sigma} (1-n_{i,-\sigma})
\,.
\end{equation}
At half--filling the $t$-$J$ Hamiltonian reduces to the antiferromagnetic
Heisenberg model.\\

Experiments show that already a few percent of additional holes
(away from half--filling) destroy the long--range antiferromagnetic order
in the CuO$_2$ planes. This demonstrates
the importance of the interplay between
antiferromagnetism and the motion of holes
in the high--T$_c$ superconductors.
To investigate the mutual influence one may start from the
$t$-$J$ Hamiltonian (\ref{TJ_MOD}). However, due to strong correlations usual
diagrammatic techniques based on Wick's theorem
can not be easily applied to (\ref{TJ_MOD}).
Neither the first nor the second part of (\ref{TJ_MOD}) is
bilinear in fermion operators, and the
creation and annihiliation operators $\hat{c}_{i\sigma}^{\dagger}$
and $\hat{c}_{i\sigma}$ do not obey
simple anticommutation relations. For this reason, non--standard
analytical methods like variational wavefunctions, coupled--cluster
methods, or slave--boson and slave--fermion
techniques\cite{Schmitt,Kane} have been employed.

In the following, we present a static approach to evaluate static
and dynamical properties
in weakly doped antiferromagnets. The calculations are based on a cumulant
method for computing the ground--state energy of correlated
electronic systems. This is in contrast to the usual approach to
dynamical properties which is based on dynamical
quantities like time-- or frequency--dependent correlation
functions. The paper is organized as follows: In Sec.~II we shall
describe the cumulant method which was recently
proposed in refs.~\cite{BeckFul88,BeckWonFul89,BeckBre90}.
In Sec.~III this formalism is applied to the motion of
a single hole generated in the ground state
of a quantum antiferromagnet at half--filling.
This problem was already investigated by a number of authors
\cite{Schmitt,Kane,BrinkRice,Nagaoka,Trugman88,ShrSig88,%
MaHo,EdBeck90,EdBeckStep90,BEW,LiuMan95}.
Our first aim is to show that our approach is able to
reproduce results known from literature, see for instance review
articles\cite{Dagotto94,Brenig95}.
Processes leading to the hole motion are described
within the concept of path operators which leads to the
well--known spin--bag picture. For the case of one hole we obtain a quasiparticle
dispersion which has minima at $(\pm\pi/2,\pm\pi/2)$ and a bandwidth of
$1.4J...1.5J$ in agreement with several analytical and numerical calculations.
Our main aim is to investigate the coupling of spins and
holes in the weakly doped regime. This is the subject of Sections IV and V.
In Sec.~IV we describe the ground state for the case of small doping.
Then we evaluate the staggered magnetization from the ground--state
energy by introducing an external field coupling to the
staggered magnetization. In Sec.~V we derive the dispersion relation for
spin waves in the doped system. We find that the spin wave energies
are strongly renormalized due to the presence of holes. The
spin--wave velocity vanishes at a critical hole concentration
of a few percent. The strong coupling between spin waves and
holes also explains the experimentally observed fast
decrease of the staggered magnetization with increasing hole doping.

\section{Cumulant approach}

Conventional treatments of systems with electronic correlations
start from the uncorrelated limit and include many--body effects
by perturbation theory. This includes
summation over classes of Feynman diagrams. For a treatment of strongly
correlated electrons one would like to proceed the other way round, i.e., by
starting from a local picture including the electronic correlations and by
expanding with respect to the hybridization interactions. In this case the
unperturbed Hamiltonian $H_0$ contains two--particle operators. This leads
to the difficulty that Wick's theorem is no longer applicable. Therefore,
usual diagrammatic techniques cannot be used for strongly correlated
electronic systems. For this reason other methods for treating correlated
electrons have been developed, e.g., slave--boson and slave--fermion
techniques\cite{Schmitt,Kane}.

An alternative approach for calculating expectation values and dynamical
correlation functions\cite{BeckFul88,BeckWonFul89,BeckBre90}
is based on the introduction of cumulants. Provided that the Hamiltonian
of the system can be split into $H=H_0+H_1$ with eigenstates and
eigenvalues of $H_0$ known, this method uses the decomposition
\begin{equation}
e^{\ -\lambda H} \,=\,
e^{\ -\lambda (H_1+L_0)}\ e^{\ -\lambda H_0}
\label{EXPH_DECOMP}
\end{equation}
which can be proven by comparing the equations of motion of either side with
respect to $\lambda$. The Liouville operator $L_0$ is a superoperator
defined by $L_0 A = [H_0,A]_{-}$.
Let us denote the ground state of the unperturbed Hamiltonian $H_0$
by $|\phi_0\rangle$ and its energy by $\epsilon_0$
\begin{equation}
H_0 |\phi_0\rangle = \epsilon_0|\phi_0\rangle .
\end{equation}
For the following it is useful to introduce a cumulant
bilinear form defined by
\begin{equation}
(A|B) = (\phi_0|A^\dagger B|\phi_0) :=
\langle\phi_0|A^\dagger B|\phi_0\rangle^c ,
\end{equation}
where $\langle...\rangle^c$ denotes a cumulant expectation value. For a
detailed discussion of cumulants see e.g. Kubo\cite{Kubo}.

Our aim is to calculate the ground state energy $E_0$ of $H$:
\begin{equation}
H|\psi_0\rangle = E_0|\psi_0\rangle .
\end{equation}
Using (\ref{EXPH_DECOMP}) one can show\cite{BeckWonFul89} that $E_0$ is
given by
\begin{eqnarray}
E_0 &=& \epsilon_0 + (H_1|\Omega) = (H|\Omega) , \label{KUMGSEN}\\
\Omega &=& 1 + \lim_{x \to 0} {1 \over {x-(L_0+H_1)}}H_1 \nonumber.
\end{eqnarray}
The operator $\Omega$ has
similarity to the so--called wave operator (or Moeller operator known from
scattering theory). It transforms the ground state $|\phi_0\rangle$ of the
unperturbed system into the exact ground state $|\psi_0\rangle$ of $H$.

The relation (\ref{KUMGSEN}) can be applied to either weakly or strongly
correlated systems because its use is independent of the operator
statistics , i.e., it is valid for fermions, bosons or spins.
Based on the same approach also dynamical correlation
functions can be calculated\cite{BeckBre90}.
For a derivation and detailed discussion of relation (\ref{KUMGSEN}) see
refs.~\cite{BeckWonFul89,Fulde}.

Treating cumulant expectation values one must distinguish between prime and
composite operators. A prime operator is a single entity in the cumulant
evaluation procedure. Expanding $\Omega$ given in (\ref{KUMGSEN})
the resulting products of $L_0$ and $H_1$
are composite operators in the cumulant ordering. Assuming that
$\Omega$ can be represented by an universal analytic function $f$ of
a prime operator $S$, i.e., $\Omega=f(S)$, it has been shown \cite{SchorkFul92}
that $\Omega$ must be of the form
$\Omega = f(S) = \exp{S}$.
As a generalization we shall use in the following an exponential form
as an ansatz for $\Omega$, i.e.,
\begin{equation}
\Omega = e^{\ \sum_\nu\lambda_\nu S_\nu}
\end{equation}
with a finite set of relevant operators $S_\nu$ replacing $S$.
The $S_\nu$ have to be chosen
in such a way that $\exp(\sum_\nu\lambda_\nu S_\nu) |\phi_0\rangle$
(with appropriate parameters $\lambda_\nu$) represents a good approximation
of the exact ground state. Following ref.~\cite{SchorkFul92}
we obtain a set of coupled equations for calculating the
ground--state energy $E_0$:
\begin{eqnarray}
E_0 &=& (H|{\rm e}^S) , \nonumber\\
0   &=& (S_{\mu}|H{\rm e}^S) , \label{KUMVER}\\
S   &=& \sum_{\nu}\lambda_{\nu}S_{\nu} \nonumber.
\end{eqnarray}
The expansion coefficients $\lambda_\nu$ can be determined from the equation
(\ref{KUMVER}$)_2$. Note that the relation $0=(A|H\Omega)$ holds for all
operators $A$ (see ref.~\cite{SchorkFul92}).

In several cases the set of non-linear equations (\ref{KUMVER}) can be transformed
into an eigenvalue problem. First, the cumulant expectation values with
the exponential can be evaluated according to appendix A. This leads to
\begin{eqnarray}
E_0 &=& \langle H {\rm e}^S \rangle , \nonumber\\
\langle S_\mu^+ H {\rm e}^S \rangle  
    &=& E_0 ~\langle S_{\mu}^+{\rm e}^S \rangle,~\mu=1,...
\label{KUMVER2}
\end{eqnarray}
with $S$ defined in (\ref{KUMVER}$)_3$.
With the formal definition $S_0=1$ and the assumption
$\langle S_\nu \rangle=0$ for all operators $S_\nu$ ($\nu=1,...$) in $S$
we can formally include (\ref{KUMVER2}$)_1$:
\begin{eqnarray}
\langle S_\mu^+ H {\rm e}^S \rangle  
    &=& E_0 ~\langle S_{\mu}^+{\rm e}^S \rangle,~\mu=0,1,...
\label{KUMVER3}
\end{eqnarray}
If we now consider the case that any products of operators $S_\nu$
can be expressed by linear combinations of other operators from the set
$\{S_\nu\}$ we can decompose the exponential into
\begin{equation}
\exp(\sum_{\nu=1}\lambda_\nu S_\nu) = \sum_{\nu=0}\beta_\nu S_\nu~.
\label{expexp1}
\end{equation}
where we have introduced a new set of parameters $\{\beta_\nu\}$.
Inserting (\ref{expexp1}) into (\ref{KUMVER3}) leads to a generalized
eigenvalue problem:
\begin{eqnarray}
\sum_{\nu=0}\beta_\nu \langle S_\mu^+ H S_\nu \rangle  
    &=& E_0 ~ \sum_{\nu=0}\beta_\nu \langle S_{\mu}^+ S_\nu \rangle,~\mu=0,1,...~.
\label{KUMVER4}
\end{eqnarray}
Note that (\ref{KUMVER3}) is a priori size--consistent even if the sum 
in $S$ is restricted to a finite
set of operators $S_\nu$ due to the exponential which contains 
perturbations of the ground state up to infinite order. Note that, however,
(\ref{KUMVER4}) with a finite set $\{S_\nu\}$ is no longer size--consistent
since the state $\sum_\nu \beta_\nu S_\nu |\phi_0\rangle$
contains only a finite number of excitations. This is the same case as in a
configuration--interaction calculation.

\section{Hole motion (revisited)}

In this section we use the present formalism to describe one hole moving in an
undoped antiferromagnet (for hole concentration $\delta \to 0$).
This problem has already been studied in a number of papers
\cite{Schmitt,Kane,BrinkRice,Nagaoka,Trugman88,ShrSig88,%
MaHo,EdBeck90,EdBeckStep90,BEW,LiuMan95}. The main contribution to the
hole motion is the following: The motion of the hole locally
destroys the antiferromagnetic spin order leading to a string
of spin defects. However, quantum spin fluctuations
can repair pairs of frustrated spins. This process
leads to a coherent motion of the hole for each of the two sublattices.
Some of the above calculations are based
on the motion in an ideal antiferromagnetic
background, e.g.~\cite{Trugman88,ShrSig88,EdBeck90}.
However, spin fluctuations in the antiferromagnetic ground state
allow for additional hole motion processes and
thus influence the properties of the hole quasiparticles
\cite{EdBeckStep90,BEW}. Several authors have also studied
the hole motion using numerical methods like exact diagonalization and
Quantum Monte Carlo methods.

In the following we discuss the hole dispersion relation by calculating
the ground--state energy of a system at half--filling
which contains one additional hole with fixed momentum $\bf k$.
Contrary to usual methods this approach is based on a static view
of the hole motion problem. For a proper description
of the one--hole states we use the concept
of path operators\cite{BrinkRice,Nagaoka,Trugman88,ShrSig88}
and define path concatenation operators $A_n=A_{n\uparrow}+
A_{n\downarrow}$. The operators
$A_{n\uparrow}$ and $A_{n\downarrow}$ refer to the two sublattices.
$A_n$ operating on the N\'{e}el state with one hole,
$\hat c_{i\uparrow} |\phi_{N{\mathaccent 19 e}el}\rangle$, moves the hole
$n$ steps away and creates a path or string of $n$ spin defects
attached to the transferred hole. Explicitely, the operators $A_{n\uparrow}$
are defined by

\begin{eqnarray}
A_{1\uparrow}\quad &=&\quad \frac{-1} {\sqrt{z_{0}}}\quad\sum_{ij}\hat{c}_{j\downarrow}
        \hat{c}_{i\downarrow}^{+}\, {\tilde R}_{ji}\, ,\nonumber\\
A_{2\uparrow}\quad &=&\quad \frac{1} {\sqrt{z_{0}\, (z_{0}-1)}}\quad\sum_{ijl}
        \hat{c}_{l\uparrow}S_{j}^{+}\hat{c}_{i\downarrow}^{+}
        \, R_{lj}^{(i)}\, {\tilde R}_{ji}\, ,\\
A_{3\uparrow}\quad &=&\quad \frac{-1} {\sqrt{z_{0}\, (z_{0}-1)^{2}}}\quad\sum_{ijlm}
        \hat{c}_{m\downarrow}S_{l}^{-}S_{j}^{+}\hat{c}_{i\downarrow}^{+}
        \, R_{ml}^{(j)}\, R_{lj}^{(i)}\, {\tilde R}_{ji}\, ,\nonumber\\
\ldots\quad&\quad&\nonumber\\
(&i&\in\uparrow,\, j\in\downarrow,\, l\in\uparrow,\, m\in\downarrow)
\nonumber
\end{eqnarray}
The operators $A_{n\downarrow}$ for the
'down' sublattice are defined analogously with all spins reversed.
$z_0$=4 denotes the number of nearest neighbor
sites in the lattice. The matrices ${\tilde R}_{ji}$
and $R_{lj}^{(i)}$ allow the hole to jump to its four
nearest neighbors in the first step and to only three
new nearest neighbors by hopping forward in each
further step:
\begin{eqnarray}
{\tilde R}_{ji}\quad &=&\quad\left\{
   \begin{array}{rl}
     1&\quad i,\, j\quad {\rm nearest\,neighbors}\\
     0&\quad {\rm otherwise}\end{array}\right.\quad ,\\
R_{lj}^{(i)}\quad &=&\quad\left\{
    \begin{array}{rl}
      1&\quad j,\, l\quad {\rm nearest\,neighbors\,and}\, l\not= i \\
      0&\quad {\rm otherwise}
    \end{array}\right.\quad . \nonumber
\end{eqnarray}

Let us split the Hamiltonian (\ref{TJ_MOD}) into an unperturbed part
$H_0$ and into a perturbation $H_1$ according to
\begin{eqnarray}
H_{0}\, &=&\quad H_{Ising}\,\, =\,\, J\,
  \sum_{<ij>}(S_{i}^{z}\, S_{j}^{z}\,-\, \frac{n_{i} n_{j}} {4}\, )
  \,+\, J\, (N-2)\,\, ,\nonumber\\
H_{1}\, &=&\quad H_{t}\, + H_{\bot} \\
&=&\quad-t\, \,\sum_{<ij>,\sigma}(\,\hat{c}_{i\sigma}^{+}\,
  \hat{c}_{j\sigma} \,+\, \hat{c}_{j\sigma}^{+}\,\hat{c}_{i\sigma}\,)
  \,+\, \frac{J} {2}\,\sum_{<ij>}(\, S_{i}^{-}S_{j}^{+}\,+\, S_{i}^{+}S_{j}^{-}
  \, )\, \nonumber.
\end{eqnarray}
The unperturbed Hamiltonian $H_0$ is essentially the Ising part of the
Heisenberg exchange in (\ref{TJ_MOD}) whereas $H_1$ contains the
transverse part as well as the conditional hopping contribution.
The ground state $|\phi_0\rangle$ of $H_0$ with one hole
with momentum ${\bf k}$ is given by
\begin{equation}
|\phi_0\rangle\, =\quad {{1} \over {\sqrt{N/2}}}\,\, \sum_{i \in \uparrow}
                 {\rm e}^{i{\bf k R}_i}\, c_{i\uparrow}\,
                 |\phi_{N{\mathaccent 19 e}el}\rangle \quad.
\label{GSDEF2}
\end{equation}
For the wave operator $\Omega$ we use the exponential ansatz of Sec.~II.
In order to include the path operators $A_n$ described above and
also to take into account ground--state spin fluctuations
generated by $H_\perp$ we choose the following form
\begin{equation}
\Omega = \exp{(\sum_{n=1}^{\infty}\lambda_n A_n)} \ \exp{(\mu A_F)}
\label{STOEROP1}
\end{equation}
Here, $A_F$ is a spin--flip operator defined by
\begin{eqnarray}
A_F   \,&=& \quad \sum_{<ij>} S_i^- S_j^+ P_{ij}\quad\quad
   (i\in\uparrow,\, j\in\downarrow) \, ,\\
P_{ij}\,&=& \quad \prod_{l(ij)} (n_{l\uparrow}+n_{l\downarrow})\,.\nonumber
\end{eqnarray}
It creates pairs of spin flips on nearest neighbor
sites and corresponds to the well--known Bartkowski
wavefunction\cite{Bartkowski} which describes spin fluctuations
in the ground state of the Heisenberg antiferromagnet alone. The
projector $P_{ij}$ prevents spin fluctuations next to the
hole ( $l(ij)$ denotes all sites next to the pair of nearest neighbor
spins $ij$ ). It is introduced to avoid ambiguities since such an
excitation is equivalent to a path of length 2 and is described
by the path operator $A_2$. Note that $A_F$ does not commute
with the path operators $A_n$ because of the presence of the
projector $P_{ij}$. Note also that we have introduced in
$\Omega$ (\ref{STOEROP1}) a product of two exponentials rather than one.
This was done for simplification of further calculations and
can be considered as an extension of (\ref{KUMVER}).
For the same reasons we shall also take into account
spin fluctuations only up to the first order in $A_F$, i.e.,
\begin{equation}
\exp(\mu\,A_F)\, \approx \,1+\mu A_F\,.
\label{EXP}
\end{equation}
Using (\ref{EXP}) we shall find $-\mu = {1 \over 6} \ll 1$ which means
that higher order terms in $A_F$ are indeed neglegible.

Following the method described in Sec.~II we obtain the following
set of non--linear equations to determine the ground--state
energy $E_0$ and the coefficients $\lambda_n$.
\begin{eqnarray}
E_{0}\, &=&\quad (H |\,
      \exp(\sum_{n=1}^{\infty}\lambda_n A_n)\,(1+\mu A_F) )\, ,\nonumber\\
0\, &=&\quad ( (1+\mu A_F)\,A_i\,|H \,
      \exp(\sum_{n=1}^{\infty}\lambda_n A_n)\,(1+\mu A_F) ) ) ,\quad
      i=1,2,...\, ,\label{KUM2} \\
0\, &=&\quad (A_F\, |H\,\,
      \exp(\sum_{n=1}^{\infty}\lambda_n A_n)\,(1+\mu A_F) )\, .\nonumber
\end{eqnarray}
Note that for algebraic reasons we have used composite operators $A$
in the second of these equations (compare (\ref{KUMVER})).
In (\ref{KUM2}) all cumulant expectation values have to be
taken with respect to the unperturbed ground state (\ref{GSDEF2}).
As described at the end of Sec.~II we use the relations given in appendix A 
to transform the set of coupled non--linear equations
(\ref{KUM2}$)_2$ into a generalized eigenvalue problem.

After evaluating the cumulants and expanding the exponential
one obtains terms with products of path
operators. For such a product of two path operators
either the two operators couple to different holes creating
two different paths, or the second path operator couples to the
first one concatenating both paths. In the latter case
both paths couple to the same hole. Therefore, their effect can be
written as one path operator with both lengths
added. Thus, after having evaluated the
cumulants we introduce new path coefficients $\beta_n$ instead of the
$\lambda_n$. Application of $\Omega$ on the unperturbed ground state
leads to
\begin{equation}
\exp{(\sum_{n=1}^{\infty}\lambda_n A_n)} \ \exp{(\mu A_F)}|\phi_0\rangle
= (1+ \sum_{n=1}^{\infty}\beta_n A_n) (1+\mu A_F)|\phi_0\rangle
\end{equation}
There is a non--linear one--to--one correspondence between the
$\beta$ and $\lambda$. In the following we only calculate
the $\beta_n$ since the knowledge of the
original $\lambda_n$ is not needed.

From (\ref{KUM2}$)_1$ we find for the energy
\begin{equation}
E_0\,= \quad (H|\sum_{n=1}^{\infty} \beta_n A_n)\,+\,\mu(H|A_F)\,.
\label{LOCHENER1B}
\end{equation}
Note that no mixed contributions in $\beta_n$ and $\mu$ are obtained because
$H$ can only repair two spin defects created by $\Omega$.
The first term in (\ref{LOCHENER1B}) depends on the hole
momentum {\bf k} and therefore describes the hole dynamics. The second
term represents the ground state
energy of a Heisenberg antiferromagnet (with one hole).
The coefficients $\mu$ and $\beta_i$ have to be determined from
(\ref{KUM2}$)_{2,3}$. By evaluating the cumulants according to Appendix A,
expanding the exponentials as described above and reintroducing
new cumulants we find
\begin{eqnarray}
&\quad&\quad (\,(1+\mu A_F)A_i|H(\sum_{n=1}^{\infty} \beta_n A_n)\, (1+\mu A_F)\,)
    \nonumber \\
&=& \quad (H|\sum_{n=1}^{\infty} \beta_n A_n) ~\times~
    (\,(1+\mu A_F)A_i| (\sum_{n=1}^{\infty} \beta_n A_n)\, (1+\mu A_F)\,)
    \label{KUM3} \\
&+& \quad \mu (A_i|\sum_{n=1}^{\infty} \beta_n A_n) ~\times~
    (A_F|H(\sum_{n=1}^{\infty} \beta_n A_n)\, (1+\mu A_F)\,) \nonumber
\end{eqnarray}
and
\begin{equation}
0\,= \quad (A_F|H(\sum_{n=1}^{\infty} \beta_n A_n)\, (1+\mu A_F)\,)
     \,+\,(A_F|H) \,+\,\mu(A_F|H A_F)\,.
\label{MYBEST}
\end{equation}
The set of equations (\ref{KUM3}) can be considered as a generalized eigenvalue
problem for the $\beta_n$ and has to be solved numerically (note
$(A_n|A_m) = \delta_{nm}$). The expectation values are calculated analogously
to a recent paper\cite{BEW}. All important contributions to the hole motion
including spiral paths\cite{Trugman88} and processes due to ground state
spin fluctuations are taken into account.
From (\ref{MYBEST}) we obtain $\mu = - {1 \over 6}$. This value can also be
derived directly from the expression for the wave operator $\Omega$ given
in (\ref{KUMGSEN}) with $H_1 = H_\perp$, see ref.~\cite{BEW}.
After having calculated the values for the coefficients
$\beta_i$ and $\mu$ we are able to evaluate the ground--state
energy $E_0$ of the system given by (\ref{LOCHENER1B}). The energy
dispersion for different values of $t/J$ is shown in Fig.~1.
This dispersion relation agrees very well
with other theoretical results and also with data found from exact
diagonalization of small clusters. The energy minima are located at
$(\pm\pi/2,\pm\pi/2)$, the total bandwidth is $1.4J...1.5J$ (see Fig.~2).

The path coefficients strongly decrease in space within a few
lattice constants, see e.g.~\cite{ShrSig88,EdBeck90,BEW}. The cloud of spin
defects surrounding the hole (spin bag) is small with an average radius of
approximately 2 lattice constants for $t/J=5$. Therefore it is
a reasonable approximation to include only the first path coefficients
to describe the quasiparticle.

The present approach can also be extended to finite hole concentrations.
If we neglect hole--hole interaction processes we arrive at a rigid--band
approximation. The Fermi surface shows the well-known hole pockets
(see Fig.~3).

Note that within the method presented here all quantities
are calculated consistently. Especially,
our results are not based on the calculations of the path coefficients
for the localized hole case\cite{ShrSig88}. This is an
improvement compared to former calculations\cite{EdBeck90,BEW}.
There the hole dispersion relation problem was tackled in
first order perturbation with respect to spin fluctuations
with a zero order wave function, i.e., with the path coefficients of the
localized hole.

\section{Ground state and staggered magnetization of the doped system}

Now we turn to the description of the spin dynamics in the doped system.
We want to focus on the coupling between spin waves and hole motion. In
the following $\delta$ denotes the hole concentration away from half--filling.
Our system with $N$ lattice sites possesses $M=\delta N$ dopant holes.

The change of the magnetic properties due to the presence of mobile holes
was already investigated within the $t-J$ model by several
authors\cite{IgFul92,BeckMusch,KhaHo,BelRi}. In some of the calculations
the so-called magnetic polaron model\cite{Kane,IgFul92,BelRi} was used
which can be derived from the $t$-$J$ Hamiltonian (\ref{TJ_MOD})
by use of slave--fermion methods. In ref.~\cite{BeckMusch} 
dynamical spin susceptibilities were calculated using projection
technique\cite{Mori,Zwanzig,Forster} instead. There
a ground state according to a rigid--band approximation
for the hole quasiparticles was used to evaluate expectation values.
However, such a ground state is not able to explain the observed strong
experimental decrease\cite{RossMig91,RossMig93} of the sublattice magnetization with
increasing hole doping.

In this section we want to determine the ground--state energy and the
staggered (sublattice) magnetization of the doped system.
In the next section we shall show how
spin--wave energies can be obtained from a ground--state calculation.
The results demonstrate that this new static approach leads to similar results
as obtained from the calculation of dynamical quantities (e.g., correlation
functions).

In the Hamiltonian we introduce an additional field $B_A$ parallel
to the $z$-axis which couples to the staggered magnetization. We use
the following decomposition:
\begin{eqnarray}
H\,\, &=&\quad H_{0}+H_{1}\, , \nonumber\\
H_{0}\, &=&\quad H_{Ising}\,+\,H_{Zeeman} \nonumber\\
&=&\quad J\,
  \sum_{<ij>}(S_{i}^{z}\, S_{j}^{z}\,-\, \frac{n_{i} n_{j}} {4}\, )
  \,+\, J\, (N-2M)\,+\,
  g_J \mu_B B_A\,(-\sum_i S_i^z\,+\,\sum_j S_j^z ) ,\\
H_{1}\, &=&\quad H_{t}\, +\, H_{\bot} \nonumber\\
&=&\quad-t\, \,\sum_{<ij>,\sigma}(\,\hat{c}_{i\sigma}^{+}\,
  \hat{c}_{j\sigma}\, \,+\, \,\hat{c}_{j\sigma}^{+}\,\hat{c}_{i\sigma}\,)
  \,+\, \frac{J} {2}\,\sum_{<ij>}(\, S_{i}^{-}S_{j}^{+}\,+\, S_{i}^{+}S_{j}^{-}
  \, )\, \nonumber.
\end{eqnarray}
The ground state $|\phi_0\rangle$ of the unperturbed Hamiltonian $H_0$ is an
antiferromagnetically ordered N\'{e}el state with $M$ holes. The holes have
fixed momenta ${\bf k}_m$ and are located on the sublattice $\sigma_m$
($\sigma_m = \uparrow,\downarrow$)
\begin{eqnarray}
|\phi_0\rangle\, &=&\quad
  \hat{c}_{{\bf k}_{1}\sigma_{1}}\, \ldots\,\hat{c}_{{\bf k}_{M}\sigma_{M}}
  \, |\phi_{N{\mathaccent 19 e}el}\rangle \nonumber\\
&=&\quad \frac{1} {(N/2)^{M/2}}\,\prod_{m=1}^{M}\,\left(
  \sum_{i_{m}\in\sigma_m}\,
  {\rm e}^{i\,{\bf k}_{m}{\bf R}_{i_m}}\,
  \hat{c}_{i_m\sigma_m}\right) |\phi_{N{\mathaccent 19 e}el}\rangle \, .
\label{GSDEF3}
\end{eqnarray}
The staggered magnetization at zero temperature can be obtained from the
ground state energy $E_0$ by
\begin{equation}
M_{eff} \,=\, \langle\psi_0|g_J\mu_B(\sum_{i \in \uparrow}  S_i^z -
                               \sum_{j \in \downarrow}S_j^z ) |\psi_0\rangle
        \,=\, - {\partial E_0 \over \partial B_A}
\label{MAG_EDIFF}
\end{equation}
where $|\psi_0\rangle$ denotes the exact ground state.

In the wave operator $\Omega$ we have to include both parts of $H_1$,
the spin-flip term $H_\perp$ and the hopping term $H_t$. Spin
fluctuations generated by $H_\perp$ can be
described by pairs of spin waves
$(S_{{\bf q}\uparrow}^- S_{-{\bf q}\downarrow}^+)$ with
\begin{eqnarray}
S_{{\bf q}\uparrow}^{-}\, &=&\quad \frac{1} {\sqrt{N/2}}\quad
  \sum_{i\in\uparrow}{\rm e}^{i{\bf q}{\bf R}_i}
  \, S_{i}^{-}\,\, , \label{SQ_DEF} \\
S_{{\bf q}\downarrow}^{+}\, &=&\quad \frac{1} {\sqrt{N/2}}\quad
  \sum_{j\in\downarrow}{\rm e}^{i{\bf q}{\bf R}_j}
  \, S_{j}^{+}\,\, . \nonumber
\end{eqnarray}
defining creation operators of magnons on both sublattices. The momenta
$\bf q$ have to be taken from the magnetic Brillouin zone.
Recently, we have shown\cite{VojBeck}
that the dynamics of the undoped system can well be described by a wave
operator containing the magnon creation operators (\ref{SQ_DEF}). This represents
an extension of the Bartkowski wavefunction\cite{Bartkowski}
used in the previous section. We include spin--flip pairs
not only on nearest neighbor sites but on sites which are
arbitrarily far away from each other.
To treat hole motion processes induced by $H_t$ we use
the path operators $A_n$ from Sec.~III without the projectors $P_{ij}$.
Both types of excitations
are included in the wave operator $\Omega$.
\begin{equation}
\Omega\,\, =\,\, {\rm exp} \left(\,
  \sum_{\bf q}\nu_{\bf q}\,
  (S_{{\bf q}\uparrow}^{-} S_{-{\bf q}\downarrow}^{+})^{\cdot}\,
  \, + \, \sum_{n}\lambda_n A_{n}\, \right)
\label{STOEROPDOT2}
\end{equation}
Note that the path operators $A_n$ commute with the spin--wave creation
operators $S_{\bf q}$ (\ref{SQ_DEF}), because they all contain only
spin--flip operators destroying the N\'{e}el
order. The dot $\cdot$ in the first term of
(\ref{STOEROPDOT2}) indicates that the quantity inside $(...)^{\cdot}$
has to be treated as a single entity in the cumulant formation.

The set of equations for the ground--state energy and the coefficients $\nu_q$
and $\lambda_n$ (analogous to (\ref{KUMVER})) reads
\begin{eqnarray}
E_{0}\, &=&\quad (H' |\Omega ) \,-\,
   {1 \over 2} N g_J \mu_B B_A (1-\delta)\, ,\nonumber\\
0\, &=&\quad ( (S_{{\bf q}\uparrow}^{-}S_{-{\bf q}
  \downarrow}^{+})^{\cdot}\, |H'\, \Omega\, ) , \label{GLSYSPFADSPINW} \\
0\, &=&\quad (\, A_{n}\, |H'\, \Omega\, ) \nonumber
\end{eqnarray}
with $\Omega$ given by (\ref{STOEROPDOT2}).
The cumulant expectation values have to be taken with respect to the unperturbed
ground state (\ref{GSDEF3}). We have shifted the energy zero level by setting
$H' = H + g_J\mu_BB_A(1-\delta)N/2$. This leads to
$\langle\phi_0|H'|\phi_0\rangle = 0$ for $M$ holes in the system which are
not on nearest neighbor sites. After evaluating the cumulants
and expanding the exponential we again obtain terms containing products
of path operators which can again be replaced by new path operators.
As discussed above we take care of path
concatenations by replacing the set of coefficients \{$\lambda_n$\} by new
coefficients \{$\beta_n$\} which have to be calculated.

To handle (\ref{GLSYSPFADSPINW}) we neglect all terms which depend
more strongly than linear on the hole
concentration $\delta$. In particular, we do not
consider hole--hole interactions which are of
order $\delta^2$. This is certainly a good approximation for
weak doping. For the energy we find
\begin{eqnarray}
(H\, |\Omega\, )\, &=&\quad \langle\,
  H_{\bot}\, \sum_{{\bf q}}\nu_{\bf q}\,
  S_{{\bf q}\uparrow}^{-}S_{-{\bf q}\downarrow}^{+} \rangle
  \, + \, \beta_{1}\, \langle\, H_{t} A_1\rangle
  \, + \, \beta_{2}\, \langle\, H_{\bot} A_2 \rangle \nonumber\\
&=&\quad \langle\phi_{N{\mathaccent 19 e}el}|H_{\bot}\, \sum_{{\bf q}}
  \nu_{\bf q}\, \, S_{{\bf q}\uparrow}^{-}S_{-{\bf q}\downarrow}^{+}\,
  |\phi_{N{\mathaccent 19 e}el}\rangle \, (1-\delta)^{2} \nonumber\\
&+&\quad\sum_{m=1}^{M}\,\left(\, \, \beta_{1}\, \langle m|H_{t}
 A_1|m\rangle\, (1-\delta) \,+\, \beta_{2}\,
 \langle m|H_{\bot} A_{2}|m\rangle\, (1-\delta)^{2}
 \right)
\label{DOTEBST}
\end{eqnarray}
where we have introduced the notation $|m\rangle$ as
abbreviation for an one--hole state:
\begin{equation}
|m\rangle = c_{{\bf k}_m \sigma_m} |\phi_{N{\mathaccent 19 e}el}\rangle .
\end{equation}
The brackets $\langle ... \rangle$ in the first line of (\ref{DOTEBST})
denote expectation values with the ground state $|\phi_0\rangle$
of $H_0$. In the second equation of (\ref{DOTEBST})
we have assumed that the holes move independently. When
calculating expectation values like $\langle\phi_0|H_\perp A_2|\phi_0\rangle$
holes which do not couple to $H_\perp$ and $A_2$ give rise to a
prefactor $(1-\delta)$ for each site where the spin operators
from both $H_\perp$ and $A_2$ act. $(1-\delta)$
describes the probability for finding a spin at such a site.
For consistency we explicitely write down
all factors $(1-\delta)$ although the actual calculation is only valid
up to first order in $\delta$. The
energy (\ref{DOTEBST}) consists of a spin-wave part and a hole part.
The latter is proportional to the
hole concentration $\delta$ (besides the factors $(1-\delta)$ mentioned above).

For the coefficients $\nu_{\bf q}$ we obtain from (\ref{GLSYSPFADSPINW}$)_2$
a set of integral equations for $\nu_{\bf q}$ when $\beta_n$ is fixed.
Eq.~(\ref{GLSYSPFADSPINW}$)_3$ is a set of coupled non--linear equations
for $\beta_n$ with fixed $\nu_{\bf q}$. Analogous to the preceeding section
it can be transformed into a generalized eigenvalue problem.
To be short, here we only state the integral equation
for the $\nu_{\bf q}$:
\begin{eqnarray}
0\quad &=&\quad ((S_{ {\bf q}\uparrow}^{-}
            S_{-{\bf q}\downarrow}^{+})^{\cdot}\, |\, H\, \Omega\, ) \nonumber\\
&=&\quad (\phi_{N{\mathaccent 19 e}el} |\, (S_{{\bf q}\uparrow}^{-}
S_{-{\bf q}\downarrow}^{+})^{\cdot+}\, H_{\perp}|\phi_{N{\mathaccent 19 e}el})\,
 (1-\delta)^{2} \nonumber\\
&+&\quad (\phi_{N{\mathaccent 19 e}el} |\, (S_{{\bf q}\uparrow}^{-}
        S_{-{\bf q}\downarrow}^{+})^{\cdot+}\,
  (H_{Ising}(1-\delta)+H_{Zeeman})\,
  \nu_{\bf q}\, (S_{{\bf q}\uparrow}^{-}S_{-{\bf q}\downarrow}^{+})^{\cdot}\,
  |\phi_{N{\mathaccent 19 e}el})\, (1-\delta)^2 \nonumber\\
&+&\quad (\phi_{N{\mathaccent 19 e}el} |(S_{{\bf q}\uparrow}^{-}
  S_{-{\bf q}\downarrow}^{+})^{\cdot+}\, H_{\bot}\, \frac{1} {2!}\, (\,
  \sum_{{\bf q}_{1}}\, \nu_{{\bf q}_{1}} \,
  (S_{{\bf q}_{1\uparrow}}^{-}S_{-{\bf q}_{1\downarrow}}^{+})^{\cdot} )^{2}\,
  |\phi_{N{\mathaccent 19 e}el})\, (1-\delta)^{4} \nonumber\\
&+&\quad\sum_{m=1}^{M}\quad (m|(S_{{\bf q}\uparrow}^{-}
  S_{-{\bf q}\downarrow}^{+})^{\cdot+} (H_{t} \beta_{1}\,
  A_{1} \,+\, (H_{Ising}(1-\delta)+H_{Zeeman}) \, \beta_{2}\, A_{2})
  \, |m)\, (1-\delta)^{2} \nonumber\\
&+&\quad\sum_{m=1}^{M}\quad (m|(S_{{\bf q}\uparrow}^{-}
  S_{-{\bf q}\downarrow}^{+})^{\cdot+} (H_{t} \beta_{1}\,
  A_{1}\,+\, H_{\bot}\, \beta_{2}\, A_{2}\, (1-\delta)\,
  )\,\sum_{{\bf q}_{1}}\nu_{{\bf q}_{1}}\, (S_{{\bf q}_{1\uparrow}}^{-}
  S_{-{\bf q}_{1\downarrow}}^{+})^{\cdot}\, |m) \, (1-\delta)^{3} \nonumber\\
&+&\quad\sum_{m=1}^{M}\quad (m|(S_{{\bf q}\uparrow}^{-}
  S_{-{\bf q}\downarrow}^{+})^{\cdot+}\,
  (H_{Ising}(1-\delta)+H_{Zeeman})\, \frac{1} {2!}\,
  (\beta_{1}\, A_{1})^{2}|m)\, (1-\delta)^{2} \nonumber\\
&+&\quad\sum_{m=1}^{M}\quad (m|(S_{{\bf q}\uparrow}^{-}
  S_{-{\bf q}\downarrow}^{+})^{\cdot+} \left(H_{t}\, \beta_{1}\,
  A_{1}\, \beta_{2}\, A_{2}\,+\, H_{\bot}\, \, \frac{1} {2!}\, (\beta_{2}\, A_{2})^{2}\, (1-\delta)\,\right)\, |m)
  \, (1-\delta)^{3} \nonumber\\
&+&\quad\sum_{m=1}^{M}\quad (m|(S_{{\bf q}\uparrow}^{-}
  S_{-{\bf q}\downarrow}^{+})^{\cdot+} H_{\perp} \frac{1} {2!}
 (\beta_{1}\, A_{1})^{2}\sum_{{\bf q}_{1}}\, \nu_{{\bf q}_{1}}\,
 (S_{{\bf q}_{1\uparrow}}^{-}S_{-{\bf q}_{1\downarrow}}^{+})^{\cdot}|m)
 \, (1-\delta)^{4}\quad .
\label{NYQQUADGL}
\end{eqnarray}
The brackets $(\phi_{N{\mathaccent 19 e}el}|...|
\phi_{N{\mathaccent 19 e}el})$ and
$(m|...|m)$ denote cumulant expectation values with
$|\phi_{N{\mathaccent 19 e}el}\rangle$ and $|m\rangle$, respectively. To
simplify the evaluation of (\ref{NYQQUADGL}) we use linear
spin--wave approximation and assume independent hole motion.
Furthermore we cut the eigenvalue problem for the coefficients $\beta_n$
after the third variable, i.e., we include only paths of lengths 0, 1, and 2.
Note that the essential process for the hole motion
is taken into account by this
approximation, i.e., a hole can hop twice via $H_t$, then the
spin defects are removed via $H_\perp$. The mechanism of hole--spin
coupling is also included: A hole hops via $H_t$, the arising spin defect
can be considered as starting point of a spin wave propagating via $H_\perp$.
With linear spin--wave approximation (\ref{NYQQUADGL}) reduces to a 
quadratic equation for each $\nu_{\bf q}$.

To find a solution of (\ref{GLSYSPFADSPINW}) (with fixed hole momenta)
we have to proceed iteratively: starting from some fixed values
for $\beta_1$ and $\beta_2$ we
first calculate the $\nu_{\bf q}$ for each $\bf q$.  The obtained values
for $\nu_{\bf q}$ are then
inserted into the eigenvalue problem for the coefficients $\beta_n$.
The solution leads to new values of $\beta_n$ for the next step.
With good initial values for $\beta_1$ and
$\beta_2$ the iteration converges after a few steps.

Taking the derivative of $E_0$ with respect to the external
field $B_A$ according to (\ref{MAG_EDIFF}) we can evaluate the doping
dependend staggered (sublattice) magnetization. Results for small
hole concentrations and different values of $t/J$ are shown in
Fig.~4. The magnetization for the undoped case becomes
$M_{eff}/M_{eff,N{\mathaccent 19 e}el} \approx 0.606$, i.e., it is
the same as in linear spin-wave theory\cite{Anderson52}. This
was expected since our approximations
are equivalent to those of linear spin--wave theory.
The magnetization decreases with increasing
hole concentration $\delta$. At a critical doping level $\delta_{c_1}$ it
vanishes indicating a magnetic phase transition to
a paramagnetic state. For instance, for $t/J=5$ the cricital concentration
$\delta_{c_1}$ is approximately 3.1 \% which is in
good agreement with experimental data.

With the coefficients $\nu_{\bf q}$ and $\beta_n$
we are able to compute static and dynamical properties of
the system. In the next section we want to show how
to determine spin--wave energies within our method.

\section{Spin-wave energies}

The ground state calculation in the preceeding section allows for the
derivation of spin--wave energies. The basic idea is to calculate the energy
of a state which is the ground state of the system with certain
fixed boundary conditions. Such boundary parameters are conserved
quantities like the electron number, the total magnetization or
the total momentum. The
ground state determined above has ($N$-$M$) electrons, zero total momentum
and zero total magnetization (if we have the same number of holes on each
sublattice).

To find an expression for the spin--wave energy we now start from an
unperturbed state $|\phi_0\rangle$ which contains one additional
spin excitation
\begin{equation}
|\phi_0\rangle\, = \, S_{{\bf K}\uparrow}^{-}\,
  \hat{c}_{{\bf k}_{1}\sigma_{1}}\, \ldots\,\hat{c}_{{\bf k}_{M}\sigma_{M}}
  \, |\phi_{N{\mathaccent 19 e}el}\rangle\,.
\label{GSDEF5}
\end{equation}
Since the total magnetization $M_{tot}$ is a conserved quantity,
\begin{eqnarray}
[{\hat M}_{tot},\, H]_- \, &=&\, 0,\quad [{\hat M}_{tot},\,
  \Omega]_-\, =\, 0\, ,\\
{\hat M}_{tot}\, &=&\, g_{J}\, \mu_{B}\, \left(\sum_{i}S_{i}^{z}
  \,+\,\sum_{j}S_{j}^{z} \right)\, ,
\end{eqnarray}
(with $\Omega$ given by the exponential form (\ref{STOEROPDOT2})~)
the perturbed state has a net magnetization of
$g_J \mu_B$ in $z$--direction and is therefore orthogonal
to the ground state calculated in the preceeding section.
A straightforward calculation based on the wave operator (\ref{STOEROPDOT2})
leads to the same equations for the coefficients $\nu_{\bf q}$ and $\beta_n$.
This is also due to our use of the spin--wave--like approximations where
the spin--wave modes are treated as independent excitations.
For the energy of the system
we find an expression which differs from the ground state energy
$E_0$ of eq. (\ref{DOTEBST}) by ($\hbar $=1):
\begin{eqnarray}
\Delta E\,=\,
\omega_{{\bf K}}\, &=&\quad 2J\, (1-\delta)
  \,+\, \frac{J({\bf K)}} {2}\, \nu_{{\bf K}}\, \, (1-\delta)^{2} \nonumber\\
&+&\quad \beta_{1}\, \,\sum_{m=1}^{M_{\uparrow}}\,
  (m|S_{{\bf K}\uparrow}^{+}\, H_{t}\, A_{1}\, S_{{\bf K}\uparrow}^{-}|m)
  \, (1-\delta) \nonumber\\
&+&\quad \beta_{2}\, \,\sum_{m=1}^{M}\, \, (m|S_{{\bf K}\uparrow}^{+}
  \, H_{\bot}\, A_{2}\, S_{{\bf K}\uparrow}^{-}|m)\, (1-\delta)^{2}
  \quad .
\label{OMEGARES1}
\end{eqnarray}
$M_\uparrow = M/2$ is the number of holes on the $\uparrow$--sublattice. The
first $m$--sum runs over all holes with $\sigma_m = \uparrow$. Note that a
spin excitation $S_{{\bf K}\downarrow}^+$ in $|\phi_0\rangle$ (\ref{GSDEF5}) would
lead to the same result because the hole momentum distribution in the ground
state should be symmetric with respect to the sublattices.
The quantities $\nu_{\bf K}$ and $\beta_n$ depending
on the hole concentration $\delta$
can be taken from the calculation in the last section.
$J({\bf q})$ denotes the Fourier--transformed exchange coupling defined by
\begin{equation}
J({\bf q}) = J z_0 \gamma({\bf q})
  = J\sum_{\bf\Delta} {\rm e}^{i\bf q\bf\Delta}
  = 2J\,(\cos {\bf q}_x + \cos{\bf q}_y) .
\end{equation}
For the undoped case ($\delta = 0$) the expression (\ref{OMEGARES1})
exactly reproduces the result of linear spin-wave theory:
\begin{equation}
\omega_{\bf K}\, =\,2J\, \sqrt{1-\gamma({\bf K})^2}\, .
\end{equation}
The spin--wave energies for different hole concentrations calculated from
(\ref{OMEGARES1}) are shown in Figs.~5 and 6.

For an analytic discussion of the ${\bf K}$-dependence of the spin--wave
energy we calculate
$\nu_{\bf K}$ for fixed path coefficients $\beta_1$ and $\beta_2$ using
approximations which are again equivalent to those of linear spin--wave theory.
(Note that the results must be inserted iteratively into the eigenvalue problem for the
$\beta_n$ to get a self--consistent solution as discussed in the last section).
Eq.~(\ref{NYQQUADGL}) is a quadratic equation for $\nu_{\bf q}$.
Its solution can be inserted into (\ref{OMEGARES1}). With
\begin{equation}
t\, A({\bf q})\, \colon = \,\sum_{m=1}^{M}
  (m|(S_{{\bf q}\uparrow}^{+}S_{-{\bf q}\downarrow}^{-})^{\cdot}
   \, H_{t} A_{1}\,
  (S_{{\bf q}\uparrow}^{-}S_{-{\bf q}\downarrow}^{+})^{\cdot}
  |m)
\end{equation}
we find
\begin{equation}
\omega_{\bf K}\, =\, 2J\, (1-\delta)
  \,\sqrt{1\,-\, \gamma({\bf K})^{2}\,-\, \frac{\beta_{1}\,
  A({\bf K)}\, \, t/J} {2}\,+\,
  \frac{(\beta_{1}\, A({\bf K})\, \, t/J)^{2}} {16}}\,.
\label{OMEGARES2}
\end{equation}
The main contribution to the renormalization of the spin--wave energy is
determined by $\beta_1$ and is therefore due to the coupling between a path
of length 1 and a spin wave. Note that terms with $\beta_2$ cancel each other.
The next process renormalizing $\omega_{\bf K}$ is expected to arise from
$\beta_3$ which we neglect here. However, $\beta_3$ is small compared to 
$\beta_1$, see Sec.~III. From this fact we expect that the error introduced
by including only paths of length 0,1, and 2 in the calculation of the
spin--hole coupling is rather small.

To perform the expectation value in $A({\bf K})$ we need to know the momentum
distribution of the hole quasiparticles. It is well established that the minimum energy
of the quasiparticle dispersion is located at $(\pm\pi/2,\pm\pi/2)$.
As a simple approximation
we neglect the effect of the Fermi surface and assume that the momenta of all holes
are $(\pm\pi/2,\pm\pi/2)$, i.e., all holes are in the centre of the hole pockets.
In the following we are only interested in small momenta $\bf K$.
The first non--trivial order of $\bf K$ leads to
\begin{equation}
\gamma({\bf K}) \,=\,\frac{J({\bf K})} {J\, z_{0}}
 \,\approx\, 1\, - \, \frac{{\bf K}^{2}} {4},
 \quad A({\bf K)}
 \approx 4\, \delta\, {\bf K}^{2} \,.
\end{equation}
Inserting this into (\ref{OMEGARES2}) we end up with the following
approximation for the spin--wave energy at small momenta $\bf K$:
\begin{equation}
\omega_{\bf K}\, =\,\sqrt{2}\, J\, |{\bf K}|
 \, (1-\delta)\, \sqrt{1 \,-\,  4 \delta\, \beta_{1}\,  t/J}\, .
\end{equation}
The ${\bf K}$--dependence remains linear even upon doping. For ${\bf K} \to 0$
the spin-wave energy $\omega_{\bf K}$ goes to zero in accordance with
Goldstone's theorem. The spin--wave velocity
$v=\omega_{\bf K}/|{\bf K}|$ ($|{\bf K}| \to 0$) decreases with
doping. At a critical hole density $\delta_{c_2}$ the spin-wave
velocity vanishes. If we assume $\beta_1$ to be independent of the hole
concentration and approximate $1-\delta \approx 1$ we obtain
\begin{eqnarray}
v\, &=&\, v_{0}\, \sqrt{1\,-\, \delta/\delta_{c_2}}\, ,\label{SPW_V} \\
\delta_{c_2}\ &=&\, \frac{J} {4\, t\, \beta_{1}}\, . \nonumber
\end{eqnarray}
$v_0$ denotes the spin--wave velocity in the undoped antiferromagnet. Using
realistic values $t/J$=5, $\beta_1$=1.25 we find $\delta_{c_2}$=4.0\%.
The doping dependence of the spin--wave velocity is in good
agreement with experiments\cite{RossMig91} as well as with
other theoretical results\cite{BeckMusch,BelRi}. Fig.~7
shows the calculated doping dependence
of $v$ for $t/J=5$. Another important result to be seen in Fig. 5 is
that the $\bf K$-range, where we find a
linear $\bf K$-dependence of $\omega_{\bf K}$, becomes smaller with
increasing hole concentration. This
is also in agreement with experimental results, see e.g.
Rossat-Mignod\cite{RossMig91}. Note that the critical concentration
$\delta_{c_1}$ where the magnetization vanishes is somewhat smaller
than $\delta_{c_2}$ evaluated here. As already mentioned for $t/J=5$
we have $\delta_{c_1}\approx 3.1\%$ whereas $\delta_{c_2}\approx
4.0\%$. This is understandable since the softening of spin--wave
excitations should first lead to the vanishing of the staggered
magnetization.

The cumulant formalism provides an alternative way to derive the expression
(\ref{OMEGARES1}) for the spin--wave energy. Introducing a transverse
wave vector dependent field in the Hamiltonian one obtains
the transverse static susceptibility by taking the second derivative
of the ground state energy $E_0$ with respect to the field.
The results for the static susceptibility
allow for the calculation of spin--wave energies. The necessary link
is given by the dynamical spin susceptibility. The calculation
leads exactly to the same expression (\ref{OMEGARES1}) for the spin-wave
energy (!). The detailed derivation has been published in a
recent paper\cite{VojBeck}.
The static susceptibility for a staggered field shows
a ${\bf K}^{-2}$--divergency for all hole concentrations. This means that the
calculation presented in Sec.~IV always describes a state
with antiferromagnetic long--range order.

\section{Conclusion}

The aim of this work was to study the hole and spin dynamics
of weakly doped antiferromagnets described by
the $t$-$J$ model. Instead of using dynamical correlation functions
our approach is based on the calculation of the ground--state
energy and contains a static view of the system.

In the first part we have rederived the quasiparticle dispersion for one hole
generated in the ground state at half--filling (Fig.~1,2).
The energy minima are found at $(\pm\pi/2,\pm\pi/2)$
in agreement with other analytical and numerical results. By neglecting
hole--hole interaction the calculation can be extended to finite doping
by use of a rigid--band approximation. For the Fermi surface (Fig.~3) we
obtain the expected picture of hole--pockets around
$(\pm\pi/2, \pm\pi/2)$.

The weakly doped system has been subject of the second part of this paper.
We have presented results for the staggered magnetization and for spin--wave
energies up to first order in $\delta$. The staggered magnetization
(Fig.~4) decreases
with doping due to spin--hole interactions. The magnetization becomes zero
at a hole concentration $\delta_{c_1}$ which indicates the disappearance of
antiferromagnetic long--range order. The spin--wave energy is found to be
also strongly renormalized with hole doping (Fig.~5 and 6).
Both effects are coupled:
Due to decreasing spin--wave energies more long--wavelength spin
fluctuations are mixed into the ground state which causes the loss of
magnetization. The spin--wave velocity has a square--root
concentration dependence (Fig.~7) and vanishes for a critical hole
concentration $\delta_{c_2}$ given by (\ref{SPW_V}). Note that we find
$\delta_{c_1} < \delta_{c_2}$ (for instance, $\delta_{c_1}=3.1\%,\,
\delta_{c_2}=4.0\%$ for $t/J=5$), i.e., the magnetization vanishes
before the spin-wave velocity becomes zero. This feature was observed
in experiments\cite{RossMig91}.

The strong renormalization of the spin--wave energies due the presence of
holes arises from the interaction of spin waves with spin fluctuations
created by hole hopping. This can be interpreted as the decay of spin
waves into particle--hole pairs. To first order in $\delta$ only the
first path coefficient $\beta_1$ contributes to the renormalized
spin--wave energy (\ref{OMEGARES2}).
The destruction of the antiferromagnetic state therefore results from
the creation and annihilation of additional spin waves by moving
holes. The strength of this coupling increases with $t/J$ because
the energy gain by hole hopping is of order $t$ while the energy loss
due to a flipped spin is of order $J$. Realistic values for $t/J$ are between
3 and 5 which explains the high effectiveness of a small hole 
concentration for the destruction of antiferromagnetism.

In principle it should also be possible to discuss the influence
of spin dynamics on the quasihole properties. However,
the approximation of truncating the eigenvalue problem for the path
coefficients after the third variable ($A_2$) causes the loss
of some features of the one--hole dispersion relation.
This approximation leads to the fact that all momenta on the
boundary of the magnetic Brillouin zone correspond
to the same energy, i.e., in this case the minima are no longer obtained
at $(\pm\pi/2,\pm\pi/2)$. It would be necessary at least to include also paths
with a length of 4. This would, however, lead to a more extensive calculation.

The present calculations are only valid within the antiferromagnetic
phase of the $t$-$J$ model at small doping concentrations.
Our starting point is the N\'{e}el state which has long-range
order. Spin fluctuations are included by a perturbational method based on
cumulants.
Note that the linear spin--wave approximation employed for calculating the
expectation values becomes questionable for vanishing sublattice magnetization.
Therefore, the actual calculation can not provide a reliable description of
the system in the vicinity of the magnetic phase transition
$(\delta \approx \delta_{c_1})$.
Especially properties of the critical point are not accessible.

In the present approach, even for higher doping
the ground state has always a broken rotational symmetry. The divergency
of the staggered susceptibility for small momenta\cite{VojBeck} $\sim
1/ {\bf K}^2$ shows that we always describe a system
with antiferromagnetic long--range order. To extend our approach to
systems with only short--range magnetic order will be the subject of
future research.

\appendix
\section{Evaluation of cumulants}

In this appendix we show how to evaluate cumulants with an exponential
ansatz for the wave operator $\Omega$, compare (\ref{KUMVER}). The following
identities are useful for transforming the set of non--linear equations
(\ref{KUM2}$)_2$ into an eigenvalue problem.
\begin{eqnarray}
\langle A{\rm e}^S \rangle^c \,&=&\, \langle A{\rm e}^S\rangle\, , \nonumber\\
\langle AB{\rm e}^S \rangle^c \,&=&\, \langle AB{\rm e}^S\rangle
  - \langle A{\rm e}^S\rangle\,\langle B{\rm e}^S\rangle\, , \nonumber\\
\langle A{\rm e}^S B \rangle^c \,&=&\, \langle A{\rm e}^S B\rangle
  - \langle A{\rm e}^S\rangle\,\langle {\rm e}^S B\rangle\, , \nonumber\\
\langle ABC{\rm e}^S \rangle^c \,&=&\, \langle ABC{\rm e}^S \rangle
  - \langle AB{\rm e}^S\rangle\,\langle C{\rm e}^S \rangle
  - \langle AC{\rm e}^S\rangle\,\langle B{\rm e}^S \rangle \nonumber\\
&\quad&\quad
  - \langle BC{\rm e}^S\rangle\,\langle A{\rm e}^S \rangle
  + 2 \langle A{\rm e}^S\rangle\,\langle B{\rm e}^S \rangle\,\langle C{\rm e}^S \rangle
  \, , \nonumber\\
.&.&.
\label{KUMEXPREL}
\end{eqnarray}
These relations hold for operators with $\langle S^k \rangle=0 \forall\,k>0$,
$\langle A\rangle=0$, $\langle B\rangle=0$ etc.
They can be proven straighforwardly by expanding the exponentials,
explicitely evaluating the cumulants and then recollecting all terms.
To show this for the second identity  we start from the definition of
cumulant expectation values\cite{Kubo} for
a product of arbitrary operators $A_i$
\begin{equation}
\langle\phi|\prod_i A_i^{n_i}|\phi\rangle^c\, = \,
\left(\prod_i \left({\partial \over \partial\lambda_i} \right)^{n_i} \right)
\ln\langle\phi|\prod_i {\rm e}^{\lambda_i A_i}|\phi\rangle\,
|_{\lambda_i=0\,\forall\,i}\,.
\end{equation}
We find
\begin{eqnarray}
\langle AB S^k \rangle^c\, &=& \, {\partial \over \partial\lambda}
  {\partial \over \partial\kappa}
  \left( {\partial \over \partial\eta } \right)^k
  \ln\langle{\rm e}^{\lambda A}{\rm e}^{\kappa B}{\rm e}^{\eta S}\rangle\,
  |_{\lambda = \kappa = \eta = 0} \nonumber \\
&=&\, \left({\partial \over \partial\eta}\right)^k
  { \langle AB{\rm e}^{\eta S}\rangle \langle {\rm e}^{\eta S}\rangle \,-\,
  \langle A{\rm e}^{\eta S}\rangle \langle B{\rm e}^{\eta S}\rangle
  \over \langle{\rm e}^{\eta S}\rangle^2 }\,
  |_{\eta=0} \nonumber \\
&=&\,\langle ABS^k\rangle \,-\, \sum_{i=0}^k \left(\stackrel{k}{i}\right)
  \langle AS^i\rangle\langle BS^{k-i}\rangle\,.
\end{eqnarray}
Here we have used $\langle S^k \rangle=0$. Inserting this transformation
in the original expression leads to
\begin{eqnarray}
\langle AB{\rm e}^S\rangle^c \,&=&\, \langle AB\rangle^c\, +\,
  \sum_{k=1}^\infty\, {1 \over {k!}} \langle AB S^k \rangle^c \nonumber\\
&=&\,\langle AB{\rm e}^S\rangle \,-\, \sum_{k=0}^\infty {1 \over {k!}}\,
  \sum_{i=0}^k \left(\stackrel{k}{i}\right)\,
  \langle AS^i\rangle\langle BS^{k-i}\rangle \nonumber\\
&=&\,\langle AB{\rm e}^S\rangle \,-\, \sum_{i=0}^\infty\,
  \sum_{j=0}^\infty\, {1 \over {i!j!}}\,
  \langle AS^i\rangle\langle BS^j\rangle
\label{CUM}
\end{eqnarray}
In the third equation we have replaced
the sum over $k$ by a sum over $j$ with $j=k-i$. Eq.~(\ref{CUM}) is the
desired result.

\begin{figure}
\epsfxsize=13.5cm
\epsfysize=9cm
\epsffile{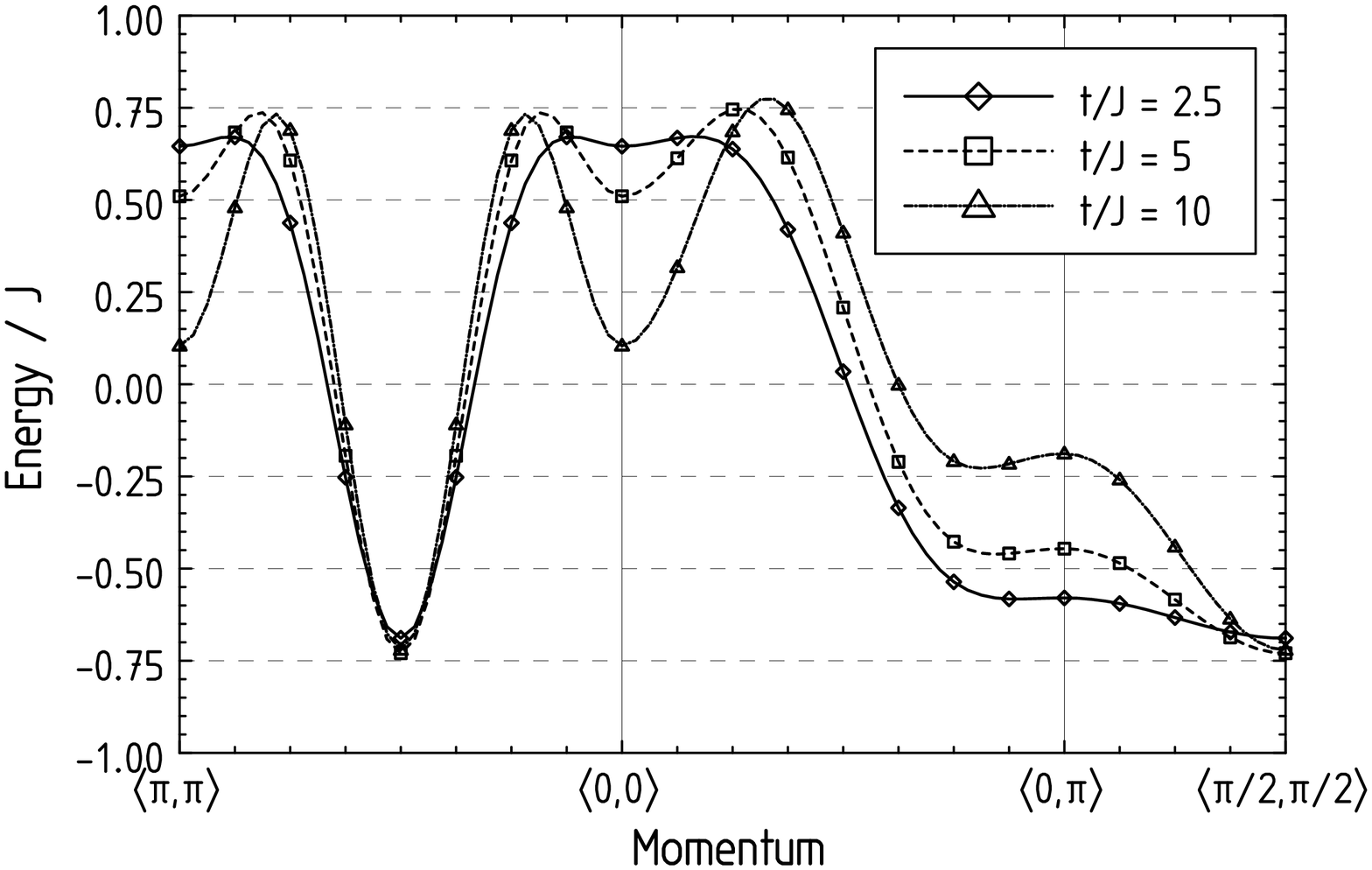}
\caption{Energy dispersion of the lowest hole excitation
for different values of
$t/J$. The zero energy was set to the center of mass of the band.}
\end{figure}

\begin{figure}
\epsfxsize=9cm
\epsfysize=9cm
\epsffile{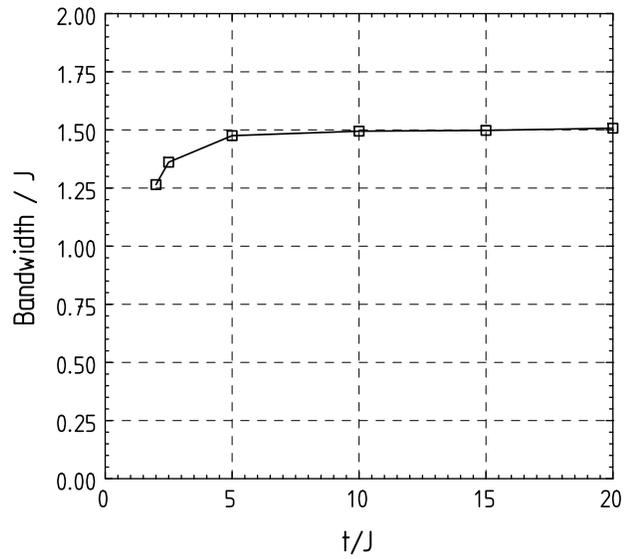}
\caption{Total bandwidth of the hole dispersion shown in Fig.1 as
function of $t/J$}
\end{figure}

\begin{figure}
\epsfxsize=11cm
\epsfysize=15.4cm
\epsffile{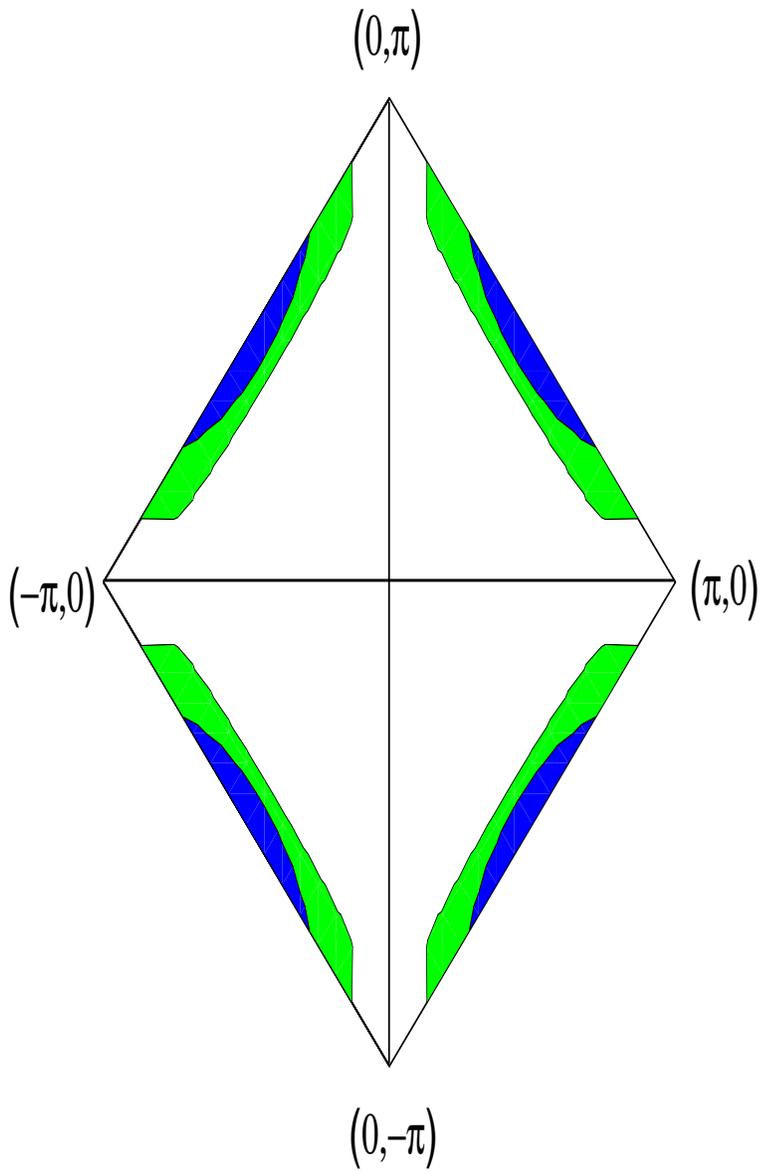}
\caption{Hole Fermi surface obtained from the dispersion relation
(Fig.~1) in a rigid band approximation for hole concentrations
$\delta$=5\% and 15\%, $t/J$=5.}
\end{figure}

\begin{figure}
\epsfxsize=14cm
\epsfysize=9.5cm
\epsffile{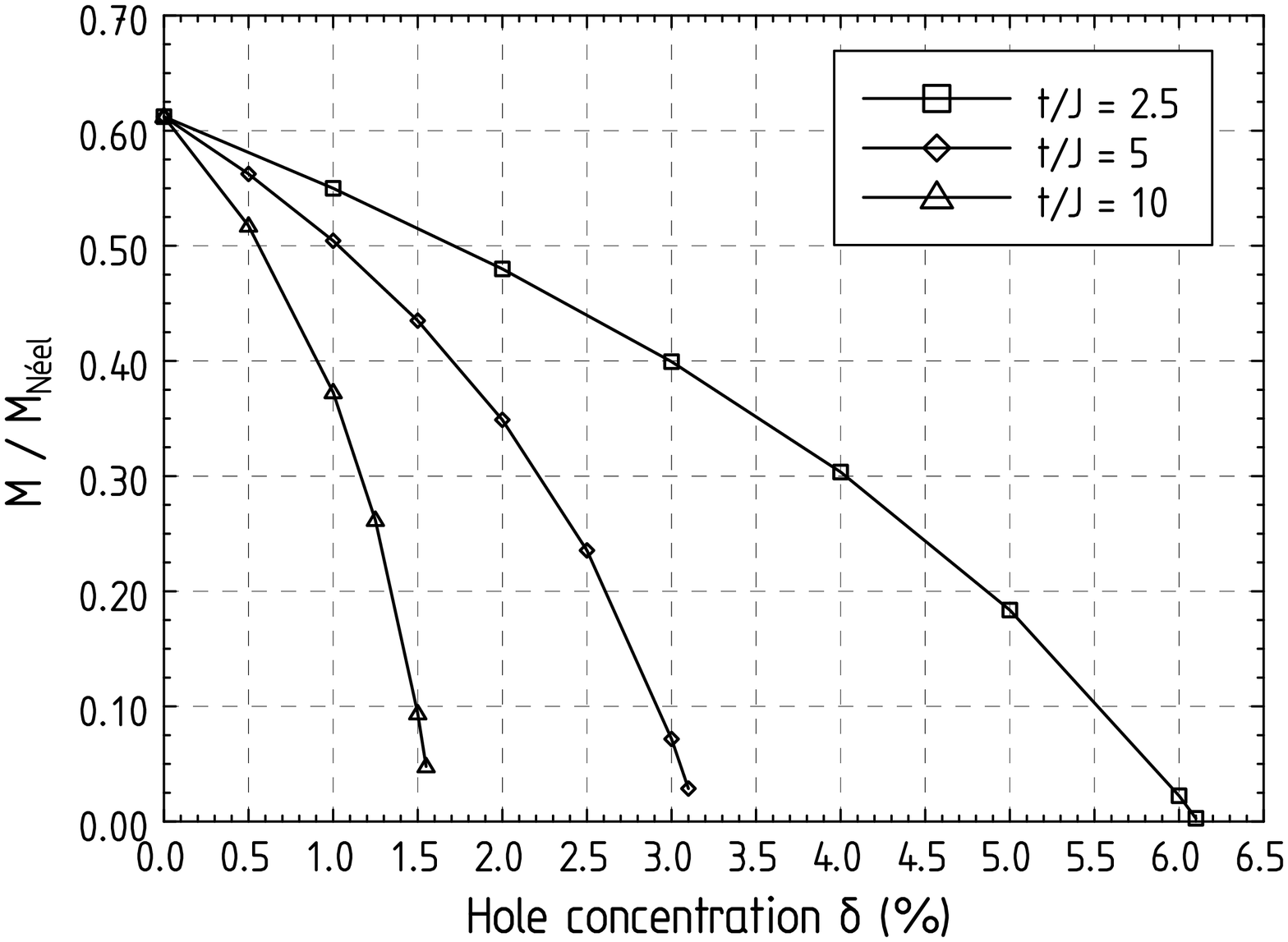}
\caption{Staggered magnetization as function of hole concentration $\delta$
for different values of $t/J$.}
\end{figure}

\begin{figure}
\epsfxsize=10cm
\epsfysize=10cm
\epsffile{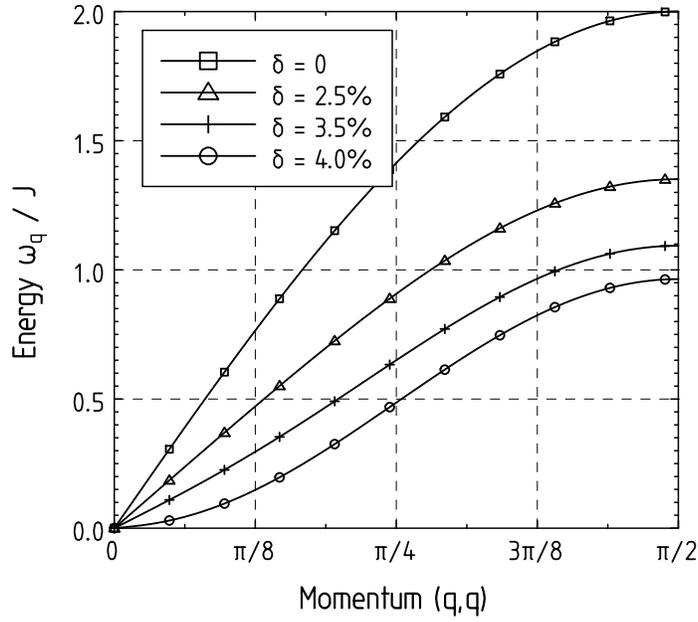}
\caption{Spin--wave energies calculated from (42) as function of momenta
$(q,q)$, for $t/J$=5 and different doping concentrations.}
\end{figure}

\begin{figure}
\epsfxsize=10cm
\epsfysize=10cm
\epsffile{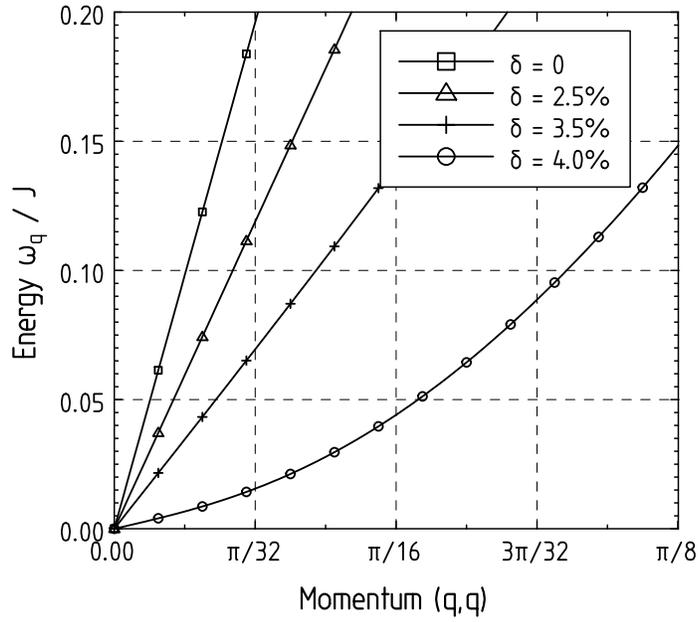}
\caption{Same as Fig.~5, but for smaller momenta $(q,q)$.}
\end{figure}

\begin{figure}
\epsfxsize=10cm
\epsfysize=10cm
\epsffile{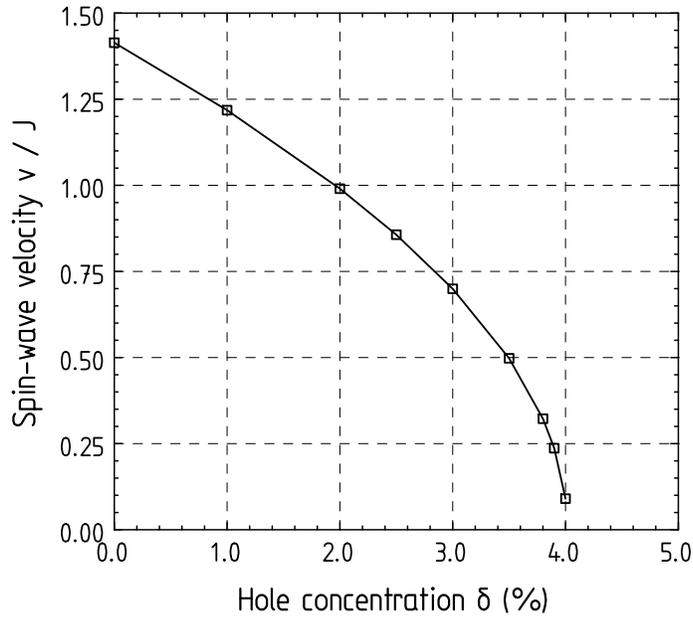}
\caption{Spin-wave velocity $v$ vs.~hole concentration $\delta$ for $t/J$=5.}
\end{figure}

\end{document}